\documentclass[sigconf]{acmart}

\usepackage{listings}
\usepackage{caption}
\usepackage{subcaption}

\AtBeginDocument{%
  }

\copyrightyear{2025}
\acmYear{2025}
\setcopyright{rightsretained}
\acmConference[SIGIR '25]{Proceedings of the 48th International ACM SIGIR Conference on Research and Development in Information Retrieval}{July 13--18, 2025}{Padua, Italy}
\acmBooktitle{Proceedings of the 48th International ACM SIGIR Conference on Research and Development in Information Retrieval (SIGIR '25), July 13--18, 2025, Padua, Italy}\acmDOI{10.1145/3726302.3730136}
\acmISBN{979-8-4007-1592-1/2025/07}

\begin{document}

\title{Navigating Speech Recording Collections with AI-Generated Illustrations}

\author{Sirina Håland}
\authornote{Both authors contributed equally to this research.}
\orcid{1234-5678-9012}
\affiliation{%
  \institution{University of Stavanger}
  \city{Stavanger}
  \country{Norway}
}

\author{Trond Karlsen Strøm}
\authornotemark[1]
\affiliation{%
  \institution{University of Stavanger}
  \city{Stavanger}
  \country{Norway}
}

\author{Petra Galu\v{s}\v{c}\'{a}kov\'{a}}
\email{petra.galuscakova@uis.no}
\affiliation{%
  \institution{University of Stavanger}
  \city{Stavanger}
  \country{Norway}
}

 \renewcommand{\shortauthors}{Sirina Håland, Trond Karlsen Strøm, \& Petra Galuščáková}

\begin{abstract}
Although the amount of available spoken content is steadily increasing, extracting information and knowledge from speech recordings remains challenging. Beyond enhancing traditional information retrieval methods such as speech search and keyword spotting, novel approaches for navigating and searching spoken content need to be explored and developed. In this paper, we propose a novel navigational method for speech archives that leverages recent advances in language and multimodal generative models. We demonstrate our approach with a Web application that organizes data into a structured format using interactive mind maps and image generation tools. The system is implemented using the TED-LIUM~3 dataset, which comprises over 2,000 speech transcripts and audio files of TED Talks. Initial user tests using a System Usability Scale (SUS) questionnaire indicate the application's potential to simplify the exploration of large speech collections.
\end{abstract}



\maketitle

\section{Introduction}
A large volume of digital spoken content has accumulated online, and diverse speech media collections such as podcasts, talks, and radio programs continue to emerge. Such speech media collections contain a wealth of knowledge and insights, but their sheer volume makes the information useless unless spoken audio can be effectively browsed and searched. Although speech recording remains one of the easiest ways to store information, acquiring information from speech collections remains problematic due to the unique properties of audio content, which is sequential and time-based.
Users are also becoming increasingly familiar with the concept of speech recordings as an information source, leading to a growing expectation that access to speech media should be as intuitive, reliable, and convenient as access to conventional text media~\citep{10.1561/1500000020}.

In this work, we aim to create a novel navigational approach for spoken collections using recent advances in generative multimodal models. Our proposed method focuses on exploratory search and zero-shot recommendation. We assume a scenario in which the user seeks to listen to recordings discussing topics of interest. We focus on a domain of informative speech recordings, such as popular science  talks and podcasts, or courses and conference recordings. A user would like to quickly review the most important topics across all available recordings, select the topics according to their interests, and listen to the recordings discussing such topics. This scenario might serve as an alternative to the more common approaches where users either select recordings purely based on metadata provided by the author of the recording or rely on recommendations based on history and popularity. Our method can lead to more serendipitous discoveries and new insights than these methods and has the potential to be more intuitive, interactive, and user-friendly than these methods. Our method is also complementary to the methods based on keywords search.

Our focus in this paper is the proposal and evaluation of this novel navigational method, which helps users explore spoken collections and discover new information. Visual representations of the content and graph-based organization of the recordings can help users gain a broader and faster understanding of the topics covered in the speech collection and assist them in finding potentially interesting recordings in a fun and interactive way. We further describe our demo implementation\footnote{\url{https://mindmap.ux.uis.no}} which uses a collection of TED talks. The source code is freely available\footnote{\url{https://github.com/SirinaHaaland/mind-map}}, and the methods can be applied to other similar collections.

\section{Related Work}
Research in Human-Computer Interaction and Information Retrieval has addressed some of the challenges in navigating speech media collections. Core approaches use Automatic Speech Recognition (ASR) to convert speech into text, enabling further analysis and retrieval. \citet{10.1561/1500000020} emphasize the need for advanced search and retrieval techniques in multimedia content, including speech media, to enhance discoverability. Their work underscores the importance of ASR as a foundational element in this domain. The TED-LIUM dataset, often used in ASR research, illustrates such efforts to enhance speech recognition accuracy \citep{ROUSSEAU14.1104, Hernandez_2018}. Research on interactive visualization techniques for multimedia content, such as the work by \citet{10.1145/1290082.1290120}, then explores how visual interfaces can support the discovery and exploration of large media collections, addressing both information overload and engagement.

Specific for podcasts, \citet{10.1145/3639701.3656324} propose a novel browsing method that uses Large Language Models (LLMs) for topic segmentation and integrates generative AI-based image cues to enhance navigation. Their study shows that segmentation and visual cues improve user comprehension and ease of navigation in podcast content compared to traditional keyword searches. Expanding on the integration of AI and multimedia, \citet{10.1145/3379337.3415845} introduce Crosscast, an automated system that leverages NLP and text mining techniques to identify geographic locations and descriptive keywords in podcast transcripts to synchronize relevant visual content with the audio narration. Their user study demonstrates that the majority of participants preferred Crosscast-generated audio-visual podcasts over traditional audio-only formats, highlighting the potential of combining audio with visuals to enhance user engagement and understanding. Building on this direction, we apply generative models to a broader range of topics.

Additionally, \citet{10.1145/3472749.3474771} present a system that generates hierarchical summaries of long-form spoken dialogues, such as interviews and podcasts, using advanced ASR and Natural Language Processing techniques. Their interface allows users to choose from varying levels of detail—short, medium, or long summaries—alongside the original ASR transcript and audio, making it easier to quickly comprehend and navigate dense spoken content according to their needs. In addition,  \citet{10.1145/3627673.3680081} focus on simplifying the structure of long recordings by automatized segmentation into chapters. In contrast to these works, we focus on visualizing the entire recording to provide an easily comprehensible overview.

Visualization of the retrieved text documents was studied by~\citet{10.1145/2484028.2484202}, who use colored distributions of the document topics. Recent advancements in video retrieval technologies are then being demonstrated in the annual Video Browser Showdown (VBS)\footnote{\url{https://videobrowsershowdown.org/}}. Tools such as VISIONE 5.0 \citep{10.1007/978-3-031-53302-0_29}, Vibro \citep{10.1007/978-3-031-53302-0_33}, PraK \citep{10.1007/978-3-031-53302-0_30}, and DiveXplore \citep{10.1007/978-3-031-53302-0_34} offer insights into how AI models and interactive systems can enhance multimedia content exploration.

\section{Interface Overview and Architecture}
We aim to address several challenges important for quality exploratory navigation in a collection of speech recordings, such as accessibility, reduced information overload, interactivity, and user engagement. To do this, we develop interactive visual graphs, referred to as Mind Maps, that guide users through the dataset. Visual representations of the content, created automatically based on the transcripts, provide a quick overview of the topic without requiring users to read or listen to the recording. This is especially important for improving accessibility and enhancing user engagement, which is particularly challenging for speech recordings.

As the system should efficiently manage and organize a vast amount of unstructured audio and transcript data without overwhelming the user, we develop and utilize a clustering algorithm to organize the audio and transcript data into coherent, manageable categories. We also implement search functionality and category filtering on these clusters to complement the exploratory interface.


\subsection{User Interface}
\begin{figure}[ht]
    \centering    \includegraphics[width=0.95\linewidth]{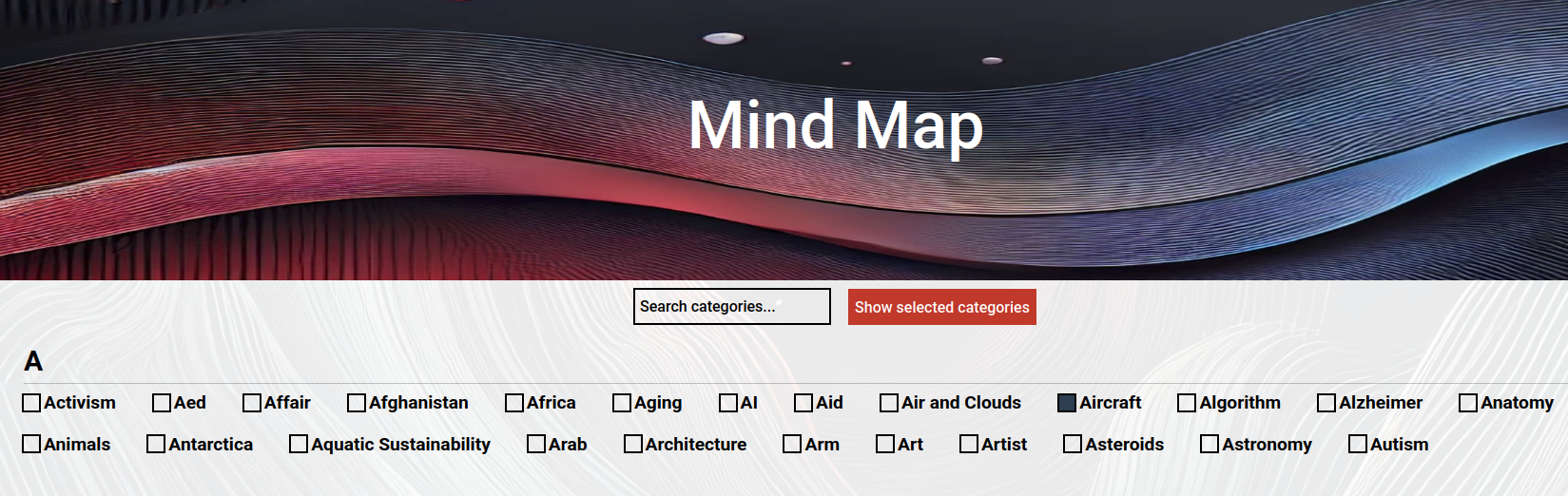}
    \caption{Layout of the Landing page with search and filtering functionality.}
    \label{fig:UIlayout}
\end{figure}

\begin{figure}[ht]
    \centering    \includegraphics[width=0.95\linewidth]{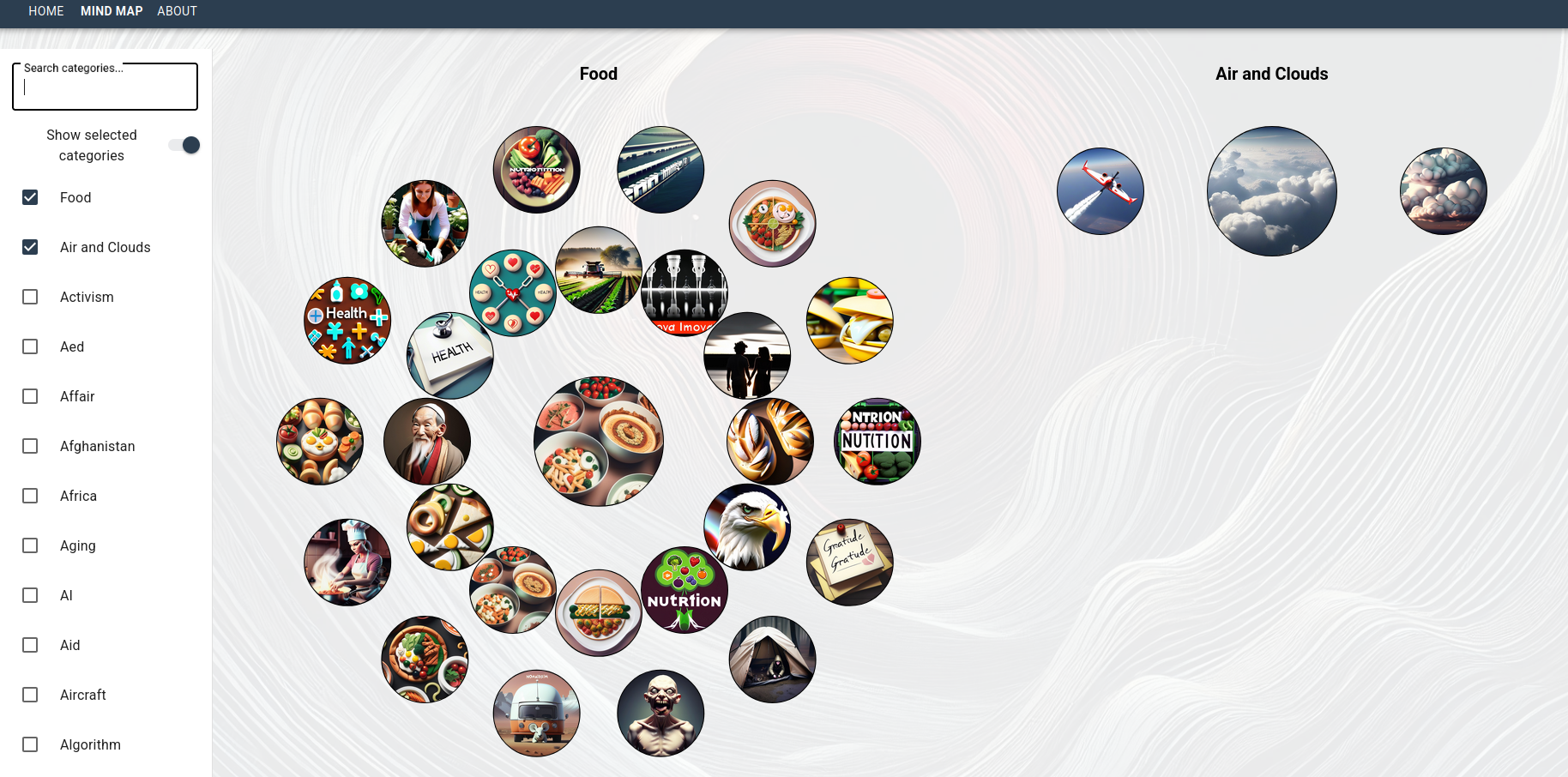}
    \caption{Example of Mind Maps (clusters of recordings based on their topical categories) on the Results page.}
    \label{fig:mainpage}
\end{figure}

The Landing page integrates search and filtering capabilities over the main topic categories discussed in the collection. All categories are displayed as interactive checkboxes on the Landing page (see Figure \ref{fig:UIlayout}), as well as in the side menu on the Results page (Figure~\ref{fig:mainpage}).

\begin{figure}[ht]
    \centering    \includegraphics[width=0.9\linewidth]{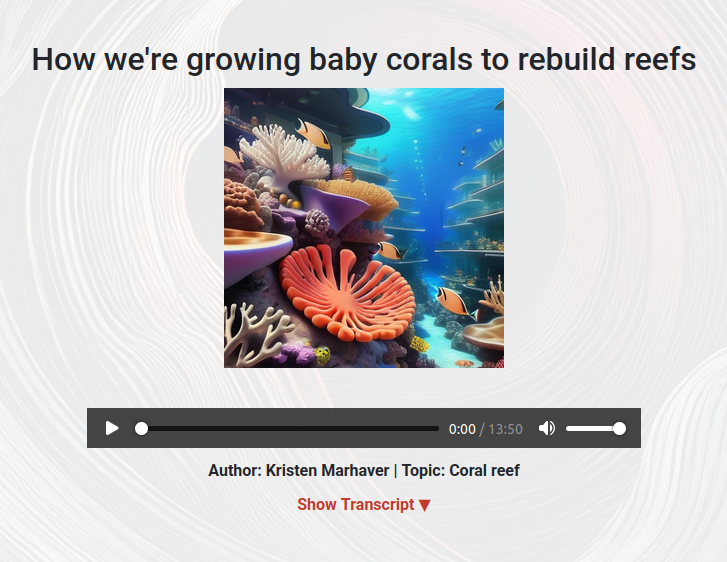}
    \caption{Layout of the page with a recording and a player.}
    \label{fig:ExMainRecPage}
\end{figure}

The main Results page, which displays selected topical categories, consists of a Mind Map (see Figure~\ref{fig:mainpage}). The goal is to create an intuitive way for users to interact with vast and extensive collections of spoken content. 
Each node in each cluster corresponds to a single recording discussing the main topic in the category and features AI-generated visuals that provide a pictorial summary of its content. The title of the recording, the name of the speaker, and a topical description of the recording are displayed on hover. Each node is a clickable link, directing users to a detailed page with audio playback, metadata, and a transcript of the recording (see Figure~\ref{fig:ExMainRecPage}). 

\subsection{Data}
We apply the interface to the TED-LIUM 3 dataset \citep{ROUSSEAU14.1104}, which is a collection of the audio recordings and their corresponding automatically created transcripts.
Audio recordings were mined from the video recordings of the TED talk. The range of topics included in the recordings is thus broad and covers subjects  such as environment, technology, and education. The collection consists of 2,351 recordings, with a mean length of 11 minutes and 30 seconds, and we assume that each recording is focused on a single main topic. The Word Error Rate of the transcripts is 6.7\%.

Raw transcripts from the TED-LIUM 3 dataset contain metadata and unknown words. 
As repetitive and frequently occurring text data such as "<NA>" and "<unk>" may skew the results of clustering, the transcripts are first preprocessed to remove metadata and unknown words. We also remove non-alphabetic characters, short words and stopwords, and lemmatize the text before applying clustering.


\subsection{Clustering}
Preprocessed transcripts are  partitioned into categories according to the main topic of the recordings. For  clustering, we initially experimented with state-of-the-art language models such as GPT-3.5 Turbo to assign a primary category to each recording. However, these models tend to generate a wide range of highly specific categories, which is suboptimal for our goal of reducing information overload, as it results in too many categories and too few recordings per category.

To avoid clusters that are either too small (leading to an excessive number of categories) or too large (containing too many recordings), we found that LLM-generated categories, as well as standard clustering methods such as BERTopic and LDA, were impractical.  Instead, we adopted a semi-automatic approach using simple statistical methods, which allowed us to control both the number and size of categories. The clustering script extends TF-IDF and K-Means clustering by incorporating user interaction. TF-IDF first converts the text data into vector representations, which are then used in K-Means clustering for categorization. A user can then interactively review the categories and corresponding file names and input the categories they wish to retain. Using this script, we created 259 categories with a relatively even distribution of recordings. Table \ref{tab:CustomPlotRes} displays the distribution of the 20 largest categories and their sizes.




\begin{table}[ht!]
\centering
\tiny
\begin{tabular}{clc}
\textbf{Order} & \textbf{Category} & \textbf{\# of Recordings} \\
\hline
1  & Computer Science  & 44  \\
2  & Climate  & 42  \\
3  & Health  & 39  \\
4  & Studying and Learning  & 36  \\
5  & Brain  & 36  \\
6  & Music  & 33  \\
7  & World  & 33  \\
8  & Technology  & 32  \\
9  & Cancer  & 28  \\
10 & Psychology & 28 \\
11 & Food & 26 \\
12 & School & 26 \\
13 & Robotics & 24 \\
14 & Linguistics & 24 \\
15 & Book & 24 \\
16 & Genetics & 23 \\
17 & Africa & 23 \\
18 & Video & 23 \\
19 & Insects & 22 \\
20 & Poetry & 21 \\
\end{tabular}
\caption{Top 20 categories generated using the interactive TF-IDF with K-Means clustering script.}
\label{tab:CustomPlotRes}
\end{table}


\subsection{Topic Illustrations}
To improve interactivity and enhance accessibility to the speech collection, we decided to use visual illustrations of the topics covered in the recordings. Such illustrations provide a quick and effective overview of the main subject of each recording. We generate illustrations for each recording and each category. After experimenting with several image generation services, we chose the image generator Novita AI\footnote{\url{https://novita.ai/}} txt2img v3 model.
Since a large number of images need to be generated, we specifically sought budget-friendly models that support API integration and produce images suitable for use as small icons in the Mind Maps. Novita AI met all these criteria, providing high-quality images at a cost of \$0.001 per image.

Our experiments with different prompt formulations showed that Novita AI performs best with short, concise descriptions. Therefore, we first summarize each recording transcript into a short description, which is then used to generate the illustration. For this, we use the GPT-3.5 Turbo model. Unlike the clustering process, which aimed for broader categories, here we prefer specific topics provided by the GPT-3.5 model. The following prompt is used to generate the topic for each transcript: \textit{``Identify the primary topic of the text in one word, similar to how `technology' might summarize a discussion on smartphones, or `environment' could describe a passage on climate change: + transcribed recording''}. As the prompt in Novita AI, we simply used the identified topic itself e.g. ``Cyber security''.

\begin{figure}[ht]
        \centering
        \includegraphics[width=\linewidth]{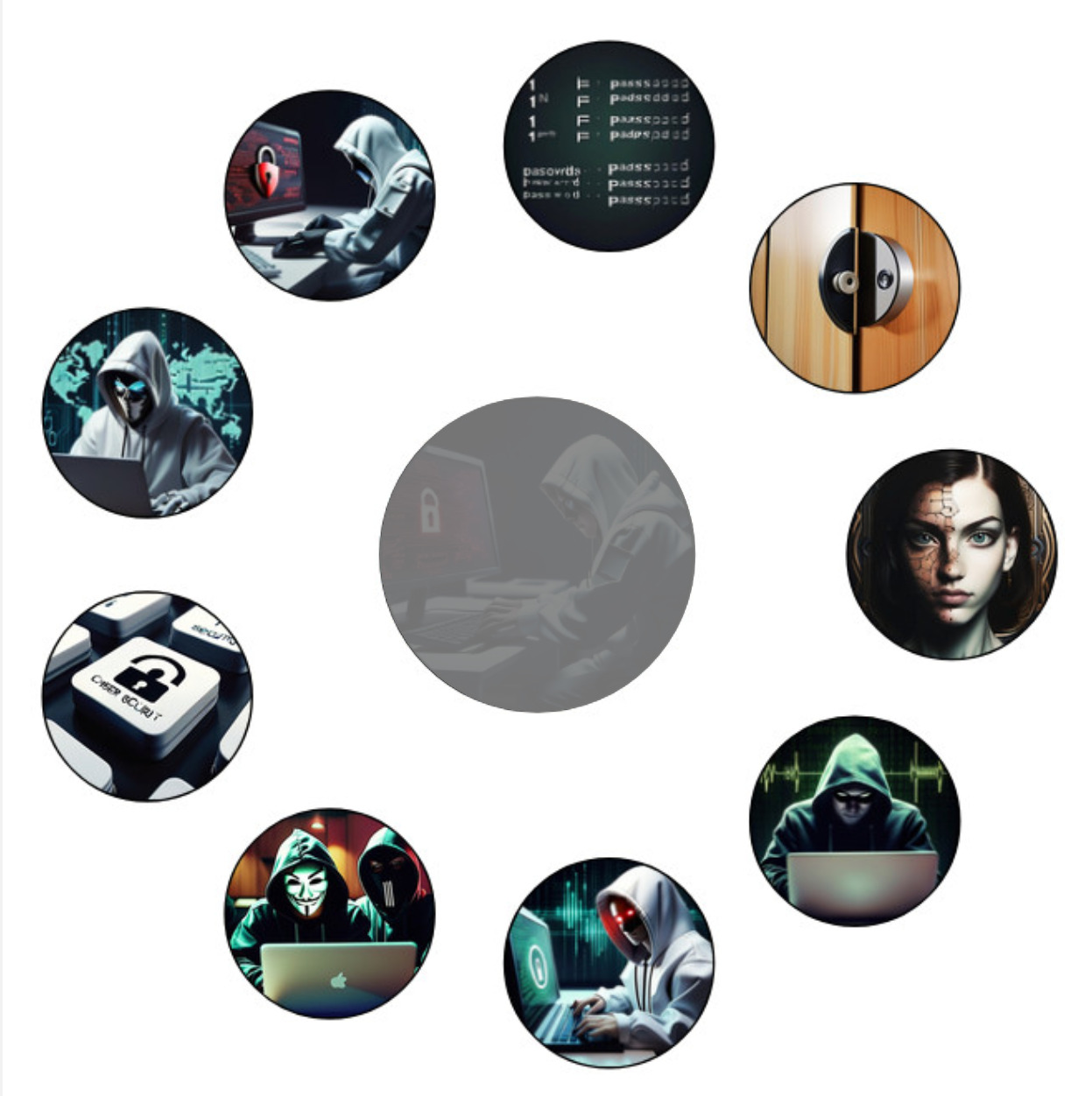} 
        \caption{Illustrations for the recordings clustered under the category Cybersecurity.}
        \label{fig:SI}
\end{figure}

This approach sometimes results in multiple recordings having the same topic. In such cases, we generate different illustrations for each recording with the same topic.
Figure~\ref{fig:SI} shows a category ``Cybersecurity''. This cluster includes recordings on the topics of ``Cyber security'', ``Passwords'', ``Hackers'', ``Hacking'', ``Privacy'' and ``Identity''.

\subsection{Implementation}
\begin{figure}[H]
    \centering    \includegraphics[width=0.7\linewidth]{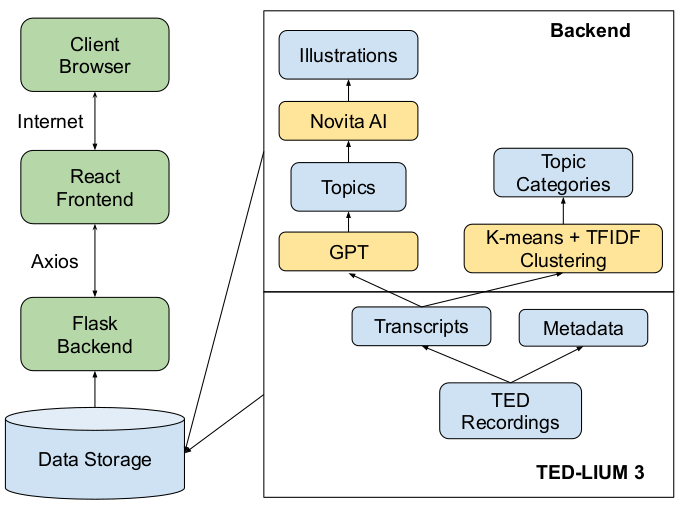}
    \caption{High-level architecture of the speech media navigation system, illustrating the interaction between components.}
    \label{fig:HLArchitecture}
\end{figure}

The demo is a single-page React application. The Axios HTTP client is used on the frontend to asynchronously fetch data from the Flask backend. The system architecture diagram (see Figure~\ref{fig:HLArchitecture}) illustrates the data flow from the web browser to the React frontend via HTTP requests, which processes the data and communicates with the Flask backend through API calls. The backend retrieves the necessary data from storage and sends it back to the frontend to update the user interface. Additionally, the backend performs preprocessing, clustering, and generates topic illustrations.

\section{Human Evaluation}
To evaluate our proposed approach, we conducted a survey based on the System Usability Scale (SUS) questionnaire. The SUS questionnaire consists of ten standardized questions that assess various aspects of system usability. Participants were asked to rate each statement on a scale from 1 (Strongly Disagree) to 5 (Strongly Agree)~\citep{Soegaard:2023:SUS}.

\begin{figure}[H]
    \centering
    \includegraphics[width=1\linewidth]{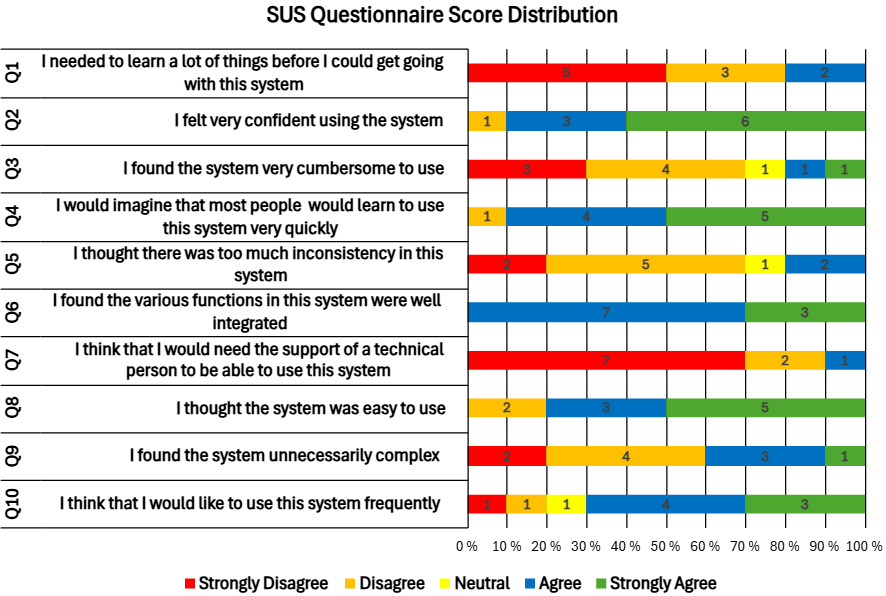}
    \caption{Distribution of user responses from the closed-ended SUS
questionnaire}
    \label{fig:susresult}
\end{figure}

The usability test involved 10 participants, including computer science students, professional programmers and developers, as well as individuals with no background in computer science. The summarized results of the SUS questionnaire are depicted in Figure~\ref{fig:susresult}.

The results from the SUS questionnaire suggest a generally positive perception of the system's usability among participants. A majority of participants found the system easy to use, with responses indicating strong agreement regarding ease of navigation, integration of functions, and user confidence. Specifically, the survey showed that most users felt confident in their ability to use the system independently without needing technical support, and they found the system’s functions well-integrated and consistent across different areas. Although most participants did not find the system complex, some users expressed concerns about its complexity and perceived cumbersomeness. These mixed responses suggest that certain aspects of the user interface could be further refined to reduce cognitive load and streamline navigation. Additionally, while the majority of users found the system easy to learn, a few participants indicated that they required more initial learning before becoming comfortable with the system. Providing initial guidelines on how to interact with the system and offering optional traditional navigational tools could facilitate faster adaptation by users and reduce cognitive load.

\section{Conclusion and Future Work}
The goal of this work was to propose and verify the concept of a novel navigational method for spoken content using automated visualization. According to the SUS questionnaire, the system’s usability is generally well-received, with strong indicators of user confidence and functional integration. However, the interface can still be further improved; for example, upgrading to a more sophisticated image generation model may enhance the descriptiveness of the images on the nodes in the Mind Maps, facilitating faster content identification by users.

A major limitation in applying this system to new collections is the semi-automatic clustering. Since the number of categories influences system usability, this part of the process is expected to benefit from human intervention, especially when applied to collections of significantly different sizes. Collections with longer and multi-topical recordings might also benefit from categorization into multiple categories and from segmenting the recordings. 

\bibliographystyle{ACM-Reference-Format}
\balance
\bibliography{sample-base}

\end{document}